\documentstyle[preprint,aps]{revtex}

\begin{document}
\title{Non-Mechanism in Quantum Oracle Computing}
\author{Giuseppe Castagnoli}
\address{Information Communication Technology Department,}
\address{Elsag spa, 16154 Genova, Italy}
\date{\today }
\maketitle

\begin{abstract}
A typical oracle problem is finding which software program is installed on a
computer, by running the computer and testing its input-output behaviour.
The program is randomly chosen from a set of programs known to the problem
solver. As well known, some oracle problems are solved more efficiently by
using quantum algorithms; this naturally implies changing the computer to
quantum, while the choice of the software program remains sharp. In order to
highlight the non-mechanistic origin of this higher efficiency, also the
uncertainty about which program is installed must be represented in a
quantum way.
\end{abstract}

\section{Introduction}

\noindent In ref. $\left[ 1\right] $, the author et al. have shown that
Simon's algorithm$^{\left[ 2\right] }$ higher than classical efficiency
comes from non-mechanism. This is a global feature of the evolution of a
quantum system undergoing wave function collapse (a revamped notion). This
evolution is driven by {\em both} the initial actions performed on the
quantum system and a final constraint imposed by the action of performing a
measurement on it.\footnote{%
The notion that there are only initial and final actions, with quantum
spontaneity (i.e. wave function collapse) in between, is due to Finkelstein$%
^{\left[ 3\right] }$.} The state produced by the initial actions should not
satisfy that constraint: in other words, having wave function collapse is
essential.\footnote{%
Quantum measurement selects {\em one} eigenstate of the measurement basis.
If the initial actions produced a superposition of such eigenstates, that
selection is a {\em final} constraint imposed on the evolution of the
quantum system. Clearly this constraint is non-redundant with the initial
actions, and is independent of the initial actions.}

The capability of driving the evolution of a quantum system by means of both
initial and final actions is leveraged in quantum computation, to obtain in
fact higher than classical efficiency. In Simon's algorithm, it can be said
that quantum measurement analogically sets a constraint on the output of a
hard-to-reverse Boolean network and, at once, solves such a constrained
network. This takes the time required to solve the {\em unconstrained}
network in all possible ways (in a quantum superposition thereof) -- namely
a time polynomial in network size. The problem of solving the constrained
network would otherwise be NP. This originates higher than classical
efficiency.

Also the efficiency of Shor's algorithm$^{\left[ 4\right] }$ is likely to
have a non-mechanistic origin, as this algorithm is tightly related to
Simon's -- to the problem of finding the period of a hard-to-reverse
function (finding the arguments given the value of the function is hard).

On the contrary, quantum algorithms which solve oracle problems reach the
solution in a deterministic way, and no wave function collapse is involved.
Apparently, this leaves no room to non-mechanism. However, as they are,
these algorithms do not capture all the relevant aspects of the problem.
Part of the problem description is not represented inside the quantum
algorithm. This can readily be corrected by introducing a small change in
the algorithm. The character of the altered algorithm is clearly
non-mechanistic. This will be shown by working on the algorithms devised by
Deutsch$^{\left[ 5,6\right] }$, Deutsch and Jozsa$^{\left[ 7\right] },$ and
Grover$^{\left[ 8\right] }$\footnote{%
Grover's algorithm is deterministic for data base size = 2 or any size in
the limit of an infinite number of loops.}.

\section{Classical and Quantum Oracles}

\noindent For our purposes, it is important to keep in mind that an oracle
problem is a competition between two parties. Sticking on Greek tradition,
let us call the examiner ``Sphinx'' and the examinee ``Oedipus''. Their
challenge is mathematically characterized as follows.

In the first place, we need to characterize the oracle. Since a computer
program establishes an input-output function, we will speak indifferently of
programs or functions. Given $B=\left\{ 0,1\right\} $, let $f_{k}$ be the $k$%
-th element of some set of functions $\left\{ f_{k}\right\}
:B^{n}\rightarrow B$, with $k=0,1,...,2^{N}-1<2^{2^{n}}$ (the latter is the
number of all possible functions from $B^{n}\ $to $B$). We say that the $k$%
-th mode of the oracle is a computer-and-program (a ``gate'') $Q_{k}$ that,
given any input $x\in B^{n}$, yields the output $f_{k}\left( x\right) \in B$%
. In other words, $k$ is an identification number of the software program
installed on the computer.

Once prepared the oracle in its $k$-th mode, the Sphinx gives it to Oedipus.
He knows all the programs in $\left\{ k\right\} $, but knows nothing about
the choice of the Sphinx. He is free to run the computer and forbidden to
inspect the mode. His problem is to identify $k$ (or a property thereof) in
the most efficient way\footnote{%
More generally, the problem is to identify $k$ with some desired probability
that such an identification is correct, but this generalization will not be
needed.}.

In quantum oracle computing, the computer becomes quantum, while its mode $k$
remains sharp (classical). This formulation cannot fit the notion of
non-mechanism. This is a global property of a quantum evolution comprising
an initial measurement (required to prepare the quantum system in a known
state), a unitary evolution and a final measurement inducing a wave function
collapse that can actively drive the evolution toward the solution of a
problem (see ref. 1).

Such an evolution should both {\em describe} and {\em solve} the problem.
For example, in Simon's and Shor's algorithms, all knowledge and ignorance
about the period of a function $f$ defined over $B^{n}$, is represented in
the superposition $\frac{1}{\sqrt{2^{n}}}\sum_{x}\left| x\right\rangle
\left| f\left( x\right) \right\rangle $. On the contrary, Oedipus' knowledge
about $\left\{ k\right\} $ and ignorance about $k$ is not represented inside
the current algorithms.

However, this can readily be adjusted by introducing a simple change in the
quantum gate. Fig. 1-a and 1-b show the usual and the altered gate -- gates
are represented in the standard form given in ref.[6]. The altered gate is
equipped with an ancillary input $k$ and computes the function $F\left(
k,x\right) $ such that, for all $k$, $F\left( k,x\right) =f_{k}\left(
x\right) $.

Oedipus' knowledge about $\left\{ k\right\} $ and ignorance about $k$ are
represented by the result of computing $F\left( k,x\right) $, with $x$
prepared in the even superposition $\frac{1}{\sqrt{2^{n}}}\sum_{x}\left|
x\right\rangle ,$ and the ancilla $k$ prepared in the even superposition $%
\frac{1}{\sqrt{2^{N}}}\sum_{k}\left| k\right\rangle $. Fig. 1-b shows the
gate $F\left( k,x\right) $ with the ancilla already prepared in that
superposition, $H$ denotes the Hadamard transform. Oedipus must be forbidden
to measure the ancilla, at least after the Hadamard transform has been
performed on it, and before solving the problem. Let us call ``gate'' the
box computing $f_{k}\left( x\right) $, or $F\left( k,x\right) $; all gate
inputs but $y$ go unchanged into the corresponding outputs, while $y$ goes
into $y\oplus f_{k}\left( x\right) $, or $y\oplus F\left( x,k\right) $,
where the sign $\oplus $ denotes module 2 addition.

\section{Quantum Algorithms for Solving Oracle Problems}

\noindent Let us consider the modified version of Deutsch's 1985 algorithm$^{%
\left[ 5\right] }$ given in ref. $\left[ 6\right] .$ $\left\{ k\right\} $ is
the set of all possible functions $f_{k}:B\rightarrow B:$

\begin{tabular}{cc}
\multicolumn{2}{c}{$k=00$} \\ 
$x$ & $f\left( x\right) $ \\ 
$0$ & $0$ \\ 
$1$ & $0$%
\end{tabular}
\qquad 
\begin{tabular}{cc}
\multicolumn{2}{c}{$k=01$} \\ 
$x$ & $f\left( x\right) $ \\ 
$0$ & $0$ \\ 
$1$ & $1$%
\end{tabular}
\qquad 
\begin{tabular}{cc}
\multicolumn{2}{c}{$k=10$} \\ 
$x$ & $f\left( x\right) $ \\ 
$0$ & $1$ \\ 
$1$ & $0$%
\end{tabular}
\qquad 
\begin{tabular}{cc}
\multicolumn{2}{c}{$k=11$} \\ 
$x$ & $f\left( x\right) $ \\ 
$0$ & $1$ \\ 
$1$ & $1$%
\end{tabular}

\noindent This set is divided into a couple of subsets: the balanced
functions, characterized by an even number of zeroes and ones and identified
by $k=01,10,$ and the unbalanced ones, which here are the constant
functions, and are identified by $k=00,11$. Oedipus' problem is to find
whether $f_{k}$ (randomly chosen by the Sphinx) is balanced or not, in the
most efficient way. Deutsch's algorithm, in one run (against two in
classical computation), deterministically yields either $\left|
1\right\rangle _{x}\left( \left| 0\right\rangle _{y}-\left| 1\right\rangle
_{y}\right) $ if the mode is balanced, or $\left| 0\right\rangle _{x}\left(
\left| 0\right\rangle _{y}-\left| 1\right\rangle _{y}\right) $ if the mode
is unbalanced. Quantum measurement in the basis $\left\{ \left|
0\right\rangle _{x},\left| 1\right\rangle _{x}\right\} $ does not induce any
wave function collapse, it just serves to read $x$: non-mechanism remains,
so to speak, hidden.

I shall now port Deutsch's algorithm to the altered gate $F$ (see fig. 2,
where an $H$ gate always denotes a Hadamard transform). The function
computed by gate $F$ is:

\begin{tabular}{cccl}
$k_{1}$ & $k_{0}$ & $x$ & $F\left( k_{1},k_{0},x\right) $ \\ 
$0$ & $0$ & $0$ & $0$ \\ 
$0$ & $0$ & $1$ & $0$ \\ 
$0$ & $1$ & $0$ & $0$ \\ 
$0$ & $1$ & $1$ & $1$ \\ 
$1$ & $0$ & $0$ & $1$ \\ 
$1$ & $0$ & $1$ & $0$ \\ 
$1$ & $1$ & $0$ & $1$ \\ 
$1$ & $1$ & $1$ & $1$%
\end{tabular}

\noindent

At this point, we need to revise the terms of the competition between the
Sphinx and Oedipus. The following protocol leaves the substance of the
problem unchanged, while adapting the problem to the algorithm given in fig.
2.

Oedipus receives the box $F\left( k,x\right) $ with $k$ prepared in the
superposition of all modes. Thus, for the time being, even the Sphinx does
not know which is the mode. Oedipus' problem is still finding whether the
mode is balanced or unbalanced in the most efficient way (without accessing $%
k$). {\em After} that Oedipus has given the solution, the Sphinx measures
register $k$, thus finding the mode; then it can check whether Oedipus'
answer was right. Of course Oedipus indirectly -- through entanglement --
affects register $k$. Things will become clearer when the protocol is
applied.

As readily checked, the output of the unitary propagation described in fig.
2 is, before measurement:

\begin{eqnarray}
\left| \varphi \left( t_{-}\right) \right\rangle &=&\left( \left|
0\right\rangle _{k_{1}}\left| 0\right\rangle _{k_{0}}-\left| 1\right\rangle
_{k_{1}}\left| 1\right\rangle _{k_{0}}\right) \left| 0\right\rangle
_{x}\left( \left| 0\right\rangle _{y}-\left| 1\right\rangle _{y}\right) + 
\nonumber \\
&&\left( \left| 0\right\rangle _{k_{1}}\left| 1\right\rangle _{k_{0}}-\left|
1\right\rangle _{k_{1}}\left| 0\right\rangle _{k_{0}}\right) \left|
1\right\rangle _{x}\left( \left| 0\right\rangle _{y}-\left| 1\right\rangle
_{y}\right)
\end{eqnarray}

Measurement of the one-qubit register $x$ induces a wave function collapse
in both registers $x$ and $k$ producing, say at time $t_{+},$ either the
outcome $x\left( t_{+}\right) =1$ or the outcome $x\left( t_{+}\right) =0.$
In the former (latter) case, register $k$ has collapsed on an even
superposition of the balanced (unbalanced) modes.

Let us call $\left| \varphi \left( t_{+}\right) \right\rangle $ the overall
registers state after that register $x$ has been measured. This state is
originated in a non-mechanistic way, being shaped by both the initial
actions, leading to $\left| \varphi \left( t_{-}\right) \right\rangle $, and
the final constraint that measurement yields a single value of $x$, which of
course means a specific value, either 1 or 0 in a mutually exclusive way:

\[
\left| \varphi \left( t_{+}\right) \right\rangle =\left| x\left(
t_{+}\right) \right\rangle _{x}\left\langle x\left( t_{+}\right) \right|
_{x}\left| \varphi \left( t_{-}\right) \right\rangle ,\text{ with }%
t_{+}>t_{-}, 
\]

\noindent see also ref.[1]. Clearly $\left| 1\right\rangle _{x}\left\langle
1\right| _{x}$ ($\left| 0\right\rangle _{x}\left\langle 0\right| _{x}$)
selects an even superposition of balanced (unbalanced) modes, see eq. (1).

Say that measurement of register $x$ yields $1$: Oedipus declares that the
mode is balanced. Then the Sphinx measures register $k$, finding either $%
k=01 $ or $k=10$ -- see eq. (1). Since either mode is balanced, the Sphinx
checks that Oedipus' answer was right.

Quantum efficiency (one run against two in classical computation) comes from
the fact that measurement of register $x$ at once {\em yields} a value of $x$
and {\em creates}, in the ancillary register $k$, an even superposition of
either balanced or unbalanced modes -- depending on $x$.

Interestingly, by backdating the outcome $k=01$ of register $k$ collapses
[ref.1,9,10] immediately after performing the Hadamard transform on the
ancilla, $k=01$ can be seen as the outcome of a random choice performed by
the Sphinx. We can see that, {\em after all} (in a literal sense), the new
protocol leaves the original problem unaltered.

State (1) shows another reason for adopting the altered algorithm. It
clearly gives the characteristic function of the balanced modes and
therefore represents all the knowledge we need to say that $\left|
1\right\rangle _{x}$ ($\left| 0\right\rangle _{x}$) means balanced
(unbalanced). On the contrary, if we started with a sharp value of $k$, the
knowledge that $x$ is such a characteristic function would necessarily dwell
in Oedipus' head; it would not be physically represented. It is reasonable
to think that, when dealing with quantum-physical computation, all the
knowledge about the object of the computation should be represented in a
physical way.

It should be evident that the foregoing considerations can be applied to the
``second'' seminal quantum oracle problem, namely Deutsch-Jozsa algorithm$^{%
\left[ 7\right] }$. Here $\left\{ f_{k}\right\} $ is the set of all the
balanced and constant functions from $B^{n}\ $to $B$. Chosen a function at
random, the problem is to efficiently find whether the quantum computer
computes a balanced or a constant function. By altering the algorithm
exactly as before, the non-mechanistic nature of the solution process
becomes manifest.

We shall now consider Grover's algorithm$^{\left[ 8\right] }$. \noindent
Given the $2^{n}$ functions $f_{k}:B^{n}\rightarrow B$ such that $%
f_{k}\left( x\right) =\delta _{k,x}$, and chosen one function $f_{k}$ at
random, Oedipus' problem is to find $k$ in the most efficient way, by
checking the input-output behaviour of a quantum computer. We shall consider
the simplest case of $n=2$, so that $k$ ranges over $00,01,10,11.$ In the
usual algorithm, $k$ is found in a deterministic way with one run of
Grover's loop. As usual, we substitute $f_{k}\left( x\right) $ with $\
F\left( k,x\right) =f_{k}\left( x\right) $ for all $k$; the two ancillary
qubits required to specify $k$ are prepared in an even superposition of all
possible modes (fig. 3).

Without entering into detail, the output of just one iteration of Grover's
loop is:

\noindent 
\begin{eqnarray*}
\left| \varphi \left( t_{-}\right) \right\rangle &=&(\left| 0\right\rangle
_{k_{1}}\left| 0\right\rangle _{k_{0}}\left| 0\right\rangle _{x_{1}}\left|
0\right\rangle _{x_{0}}+\left| 0\right\rangle _{k_{1}}\left| 1\right\rangle
_{k_{0}}\left| 0\right\rangle _{x_{1}}\left| 1\right\rangle _{x_{0}}+ \\
&&\left| 1\right\rangle _{k_{1}}\left| 0\right\rangle _{k_{0}}\left|
1\right\rangle _{x_{1}}\left| 0\right\rangle _{x_{0}}+\left| 1\right\rangle
_{k_{1}}\left| 1\right\rangle _{k_{0}}\left| 1\right\rangle _{x_{1}}\left|
1\right\rangle _{x_{0}})\left( \left| 0\right\rangle _{y}-\left|
1\right\rangle _{y}\right)
\end{eqnarray*}

\noindent Measuring $x_{1}$ and $x_{0}$, at once (non-mechanistically, as
before) {\em creates} the mode $k$ in register $k$ and yields the value of $%
k $ in register $x$. Backdating register $k$ collapse immediately after
performing the Hadarmard transform on it, yields the random choice
originally performed by the Sphinx.

\section{Conclusions}

\noindent In ref.[1] we have given a demonstration that the higher than
classical efficiency of Simon's and related algorithms comes from quantum
non-mechanism. This demonstration has been extended in this work to quantum
oracle problem solving. In conclusion, the special efficiency of all known
quantum algorithms would have a non-mechanistic origin. Non-mechanism hinges
on the notion of wave function collapse, an exclusively quantum feature that
allows to drive the evolution of a quantum system by acting on both initial
and final conditions. It is hoped that the results obtained will revamp the
notion of collapse. It seems difficult to understand the motivation of
getting rid of the notion of something that yields effective benefits%
\footnote{%
In the many universes' interpretation, collapse can be substituted by a
unitary evolution that entangles the observer with the universe. Such an
interpretation is not in contrast with the current work, its {\em %
motivations }(or a part thereof) seem to be in contrast.}.

Perhaps quantum computation, because of its unique feature of joining a
fundamental character and a capability of describing complex states of
affairs, can yield fresh insights for the interpretation of quantum
mechanics.

This work has brought in a seemingly interesting side-effect. In order to
show the non-mechanistic character of quantum oracle problem solving, a {\em %
state of knowledge} of the problem solver (Oedipus' uncertainty about the
value of $k$) had to be translated into a physical description, and this
description had to be {\em strictly quantum }in order to match reality.

Thanks are due to A. Ekert,\ D. Monti, and V. Vedral for useful discussions
and feedback.

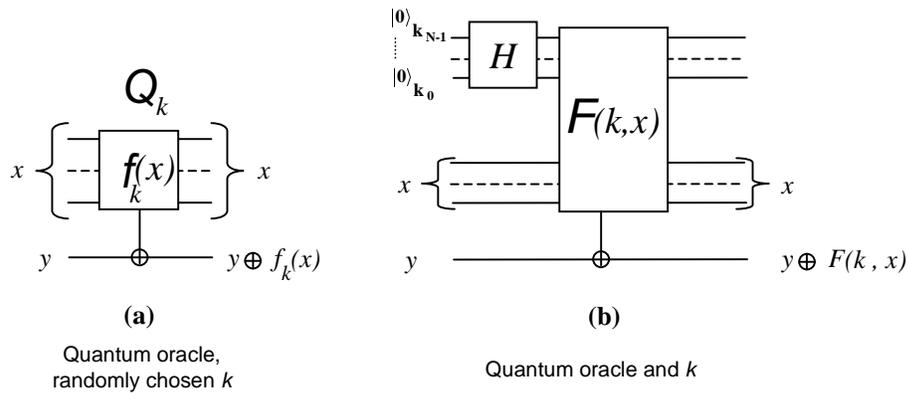
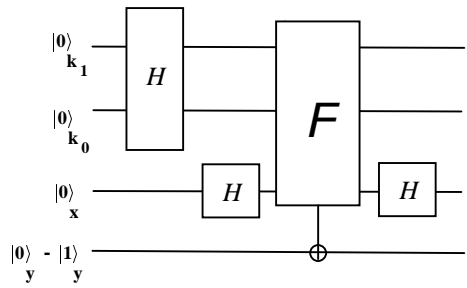
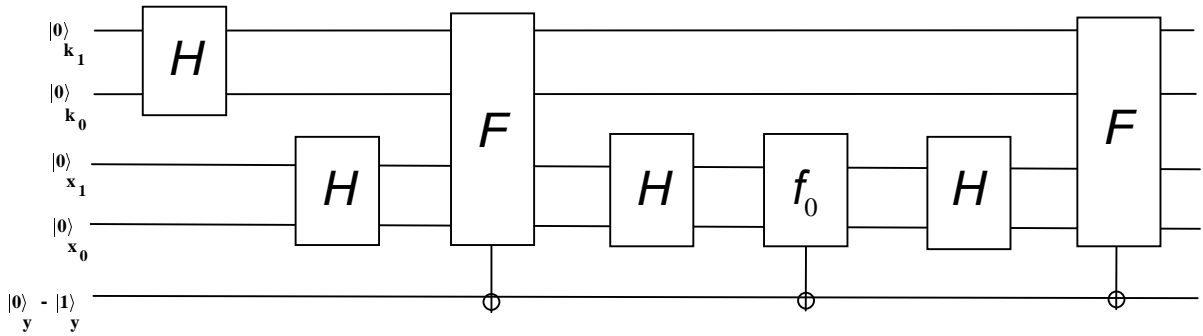
\end{document}